\documentclass{article}

\begin{document}

\title{Stable two-dimensional solitary pulses in linearly coupled dissipative
Kadomtsev-Petviashvili equations}

\bigskip

\begin{center}
Bao-Feng Feng \\
Department of Mathematics, the University of Kansas, \\
Lawrence, KS 66045, U.S.A. \\

Boris A. Malomed \\
Department of Interdisciplinary Studies, Faculty of Engineering, \\
Tel Aviv University, Tel Aviv 69978, Israel\\

Takuji Kawahara \\
Department of Aeronautics and Astronautics, \\
Graduate School of Engineering,\\
Kyoto University, Sakyo-ku, Kyoto 606-8501, Japan
\end{center}

\maketitle

\begin{center}
\textbf{Abstract}
\end{center}

We present a two-dimensional (2D) generalization of the stabilized Kuramoto
-- Sivashinsky (KS) system, based on the Kadomtsev-Petviashvili (KP)
equation including dissipation of the generic (Newell -- Whitehead -- Segel,
NWS) type and gain. The system directly applies to the description of
gravity-capillary waves on the surface of a liquid layer flowing down an
inclined plane, with a surfactant diffusing along the layer's surface.
Actually, the model is quite general, offering a simple way to stabilize
nonlinear media combining the weakly-2D dispersion of the KP type with gain
and NWS dissipation. Other applications are internal waves in multilayer
fluids flowing down the inclined plane, double-front flames in gaseous
mixtures, etc. Parallel to this weakly 2D model, we also introduce and study
a semi-phenomenological one, whose dissipative terms are isotropic, rather
than of the NWS type, in order to check if qualitative results are sensitive
to the exact form of the lossy terms. The models include an additional
linear equation of the advection-diffusion type, linearly coupled to the
main KP-NWS equation. The extra equation provides for stability of the zero
background in the system, thus opening a way to the existence of stable
localized pulses. We focus on the most interesting case, when the dispersive
part of the system is of the KP-I type, which corresponds, e.g., to the
capillary waves, and makes the existence of completely localized 2D pulses
possible. Treating the losses and gain as small perturbations and making use
of the balance equation for the field momentum, we find that the equilibrium
between the gain and losses may select two steady-state solitons from their
continuous family existing in the absence of the dissipative terms (the
latter family is found in an exact analytical form, and is demonstrated
numerically to be stable). The selected soliton with the larger amplitude is
expected to be stable. Direct simulations completely corroborate the
analytical predictions, for both the physical and phenomenological models.


\newpage

\section{Introduction and derivation of the model}

Localized structures, such as solitary pulses (SPs), play a dominant role in
many conservative and dissipative nonlinear physical systems. As is commonly
known, in conservative systems SPs are supported by balance between
nonlinearity and dispersion \cite{Ablowitz}, while in dissipative models,
such as the Ginzburg-Landau equations, it must be supplemented by the
balance between losses and gain \cite{CH}.

An important example of a one-dimensional (1D) model that combines
conservative and dissipative effects is a mixed Kuramoto - Sivashinsky (KS)
-- Korteweg - de Vries (KdV) equation, which was first introduced by Benney 
\cite{Benney} and is therefore also called the Benney equation. This
equation finds various applications in plasma physics, hydrodynamics and
other fields \cite{KSKdV,RMP}. SPs are, obviously, important objects in
systems of this type \cite{SP}; however, they cannot be completely stable
objects in the Benney equation, as the zero solution, which is a background
on top of which SPs are to be found, is linearly unstable in this equation
due to the presence of the linear gain (however, if the dispersion part of
the Benney equation is large enough, the growing perturbation, moving at its
group velocity, does not actually overlap with the SP and therefore does not
destroy it, see Refs. \cite{Chang} and references therein). A stabilized
version of the Benney equation was recently proposed in Ref. \cite{Malomed01}.
It is based on the KS -- KdV equation for a real wave field $u(x,t)$,
which is linearly coupled to an additional linear equation of the diffusion
type for an extra real field $v(x,t)$, that provides for the stabilization
of the zero background: 
\begin{eqnarray}
u_{t}+uu_{x}+u_{xxx}-v_{x} &=&-\alpha u_{xx}-\gamma u_{xxxx},  \label{u} \\
v_{t}+cv_{x}-u_{x} &=&\Gamma v_{xx}.  \label{v}
\end{eqnarray}
Here, $\alpha $, $\gamma $ and $\Gamma $ are coefficients accounting for the
gain and loss in the $u$--subsystem and loss in the $v$--subsystem,
respectively, and $c$ is a group-velocity mismatch between the fields.

It was shown both analytically (by means of the perturbation theory) and
numerically in Ref. \cite{Malomed01} that the system (\ref{u}), (\ref{v})
gives rise to a completely stable SP, as well as to stable bound states of
SPs, in a broad parametric region. As a matter of fact, Eqs. (\ref{u}) and 
(\ref{v}) furnish the first example of a model of the KS type that gives rise
to fully stable pulses, and they can be easily observed in experiment (see a
detailed discussion of physical realizations of the model -- first of all,
in terms of liquid films flowing down under the action of gravity -- in the
next section).

The liquid films and other systems in which the observation of stable SPs is
expected (e.g., double-front flames, see below) are two-dimensional (2D)
media, therefore a relevant issue is to introduce a physically meaningful 2D
version of equations (\ref{u}) and (\ref{v}) and seek for stable 2D pulses
in the generalized model. This is the objective of the present work. It will
be demonstrated that the 2D model which will be derived here is a generic
one for a number of different applications. The results will directly point
at a new type of 2D pulses that can be observed experimentally in a
straightforward way. Besides that, a possibility of the existence of stable
2D pulses may help to understand the phenomenon of turbulent spots, see,
e.g., Refs. \cite{spot} and references therein.

The paper is organized as follows. The 2D model is derived in detail in
section 2, starting with a particular physical problem, viz., a downflowing
liquid film carrying a surfactant, and then proceeding to the generic
character of the model. Parallel to the derived model, we will also consider
its counterpart which differs by the form of 2D dissipative terms, in order
to demonstrate that basic results are insensitive to particular features of
the model (which is relevant to show, even if the model is generic). In
section 3, we consider stability of the zero solution in the 2D model,
which, as well in the 1D case, is a necessary condition for the full
stability of SPs. In section 4, an analytical perturbation theory for SPs is
developed by treating the gain and loss constants as small parameters. To
this end, first a family of exact 2D soliton solutions of the zero-order
system (the one without the gain and loss terms) is obtained, following the
pattern of the well-known ``lump'' solitons of the KP-I equation. Then,
using the balance equation for the net field momentum, similar to how it was
done in the 1D model \cite{Malomed01}, we demonstrate that the combination
of the gain and loss terms may select two (or none) stationary pulses out of
the continuous soliton family existing in the zero-order system; in the case
when two stationary pulses are found, it is very plausible that the one with
the larger value of its amplitude is stable. In section 5, we present
results of direct numerical simulations of the full 2D model, which
completely confirm the analytical predictions, i.e., the existence of stable
2D localized SPs. The paper is concluded by section 6.

\section{The model}

The physical meaning of the model based on Eqs. (\ref{u}) and (\ref{v}), and
its 2D generalization developed below, can be understood in terms of a
particular application to a thin downflowing liquid layer with a surfactant
trapped on its surface. As is well established (see a review \cite{RMP}), in
the 1D case the evolution of the flow velocity field $v(x,t)$ in the layer
is governed, in appropriately chosen units, by the KS--KdV (Benney)
equation, $u_{t}+uu_{x}+u_{xxx}=-\alpha u_{xx}-\gamma u_{xxxx}$. In this
equation, the gain $\alpha $ is induced by gravity, and the loss parameter 
$\gamma $ is proportional to the fluid's viscosity coefficient, while the
left-hand side is generated by the Euler's equations (for an irrotational
flow), exactly the same way as in the classical derivation of the KdV
equation. If the surfactant is distributed on the surface of the layer with
a density $c+v(x,t)$, where $c$ and $v$ are, respectively, its constant and
small variable parts ($\left| v\right| \ll c$), the gradient of $v$ creates,
through the variation of the surface tension, an additional force $\sim
v_{x} $ which drives the flow, hence the Euler's equation adds the term 
$v_{x}$ to the right-hand side of the KS--KdV equation (the coefficient in
front of this term may be scaled to be $1$), so that the equation takes
precisely the form (\ref{u}). Further, the evolution of the surfactant
density is governed by an obvious advection-diffusion equation: $v_{t}+\left[
u\left( c+v\right) \right] _{x}=\Gamma v_{xx}$, where $\Gamma $ is the
surface diffusion constant. With regard to the condition $\left| v\right|
\ll c$, the latter equation takes the form (\ref{v}).

In the 2D case, we consider a quasi-1D (weakly 2D) flow of the film, with
the $y$--scale much larger than that along the $x$ axis. In other words, if
the wave is taken as $\exp \left( iKx+iQy\right) $, the small wavenumbers
are ordered so that 
\begin{equation}
Q\sim K^{2}.  \label{order}
\end{equation}
Then, according to the classical derivation \cite{Petviashvili70}, the KdV
part of Eq. (\ref{u}) is replaced either by the KP-I equation, 
\begin{equation}
\left( u_{t}+uu_{x}+u_{xxx}\right) _{x}=u_{yy}\,,  \label{KP-I}
\end{equation}
or by the KP-II equation, which is 
\begin{equation}
\left( u_{t}+uu_{x}+u_{xxx}\right) _{x}=-u_{yy}\,,  \label{KP-II}
\end{equation}
the coordinate $y$ being properly rescaled. The choice between Eqs. (\ref
{KP-I}) and (\ref{KP-II}) is determined by the sign of the 2D correction to
the dispersion; in particular, the capillarity gives rise to KP-I. The
difference between the KP-I and KP-II equations is that, although both of
them have quasi-1D (i.e., $y$-independent) soliton solutions that reduce to
the usual KdV solitons, only in the KP-II equation this soliton is stable
against $y$-dependent perturbations. On the other hand, the KP-I equation
has stable 2D solitons (the so-called ``lumps''), which are weakly
(non-exponentially) localized in both $x$ and $y$, see below; the KP-II
equation does not have 2D solitons.

The next step is to accordingly generalize, for the 2D situation, the
dissipative term in Eq. (\ref{u}). Dissipative generalizations of the KP
equations were introduced in some works, see, e.g., Refs. \cite{Velarde94}.
Those generalizations follow the pattern of the arrangement of the KP
equations proper: if one starts from a corresponding 1D equation containing
dissipative terms [for instance, Eq. (\ref{u})], which is written as 
$\mathrm{something}=0$, its 2D counterpart is $\left( \mathrm{something}
\right) _{x}=\pm u_{yy}$. The accordingly modified Eqs. (\ref{u}) and (\ref
{v}) then take the form 
\begin{eqnarray}
\left( u_{t}+uu_{x}+u_{xxx}-v_{x}+\alpha u_{xx}+\gamma u_{xxxx}\right) _{x}
&=&\pm u_{yy},  \label{uKP-I} \\
\left( v_{t}+cv_{x}-u_{x}-\Gamma v_{xx}\right) _{x} &=&\pm v_{yy}.
\label{vKP-I}
\end{eqnarray}
The second equation in this system can be simplified, as the usual ordering
of the partial derivatives, adopted in course of the quasi-1D derivation 
\cite{Petviashvili70}, implies that $v_{yy}$ is a small quantity of a higher
order than $v_{xxx}$, hence $v_{yy}$ may be dropped, so that Eq. (\ref{vKP-I})
remains one-dimensional. As a result, this version of the 2D system takes
the form 
\begin{eqnarray}
\left( u_{t}+uu_{x}+u_{xxx}-v_{x}+\alpha u_{xx}+\gamma u_{xxxx}\right) _{x}
&=&\pm u_{yy},  \label{usimplest} \\
v_{t}+cv_{x}-u_{x}-\Gamma v_{xx} &=&0.  \label{vsimplest}
\end{eqnarray}

However, while keeping Eq. (\ref{vsimplest}) in the 1D form is quite
acceptable, the way the dissipative and gain terms in Eq. (\ref{usimplest})
were made two-dimensional was formal, being not based on any physical
argument. Moreover, it will be shown in the next section that, unlike its 1D
counterpart (\ref{u}) and (\ref{v}), this 2D model cannot produce any stable
solitary pulses, as its zero solution is always unstable.

In order to derive a physically relevant form of the dissipative part of the 
$u$--equation, one should resort to the standard procedure that derives a 
\emph{generic} set of dissipative terms in the quasi-1D situation (in the
context of convective flows) in the Newell-Whitehead-Segel (NWS) equation 
\cite{NWS}. This equation gives a simple prescription, which, as well as the
KP equations, is based on the ordering (\ref{order}) of the longitudinal and
transverse wavenumbers: the longitudinal dissipative term $\gamma u_{xxxx}$
must be supplemented by its transverse counterpart $\gamma u_{yy}$ [the
scaling of the transverse coordinate $y$ in the NWS equation is precisely
the same which cast the KP equation in the standard form (\ref{KP-I}) or 
(\ref{KP-II}); the identity of the scalings is not accidental, being a
consequence of the fact that both KP and NWS equations are generic ones in
the quasi-1D geometry, the former one in the class of dispersive equations,
and the latter among dissipative equations]. Thus, the proper form of the 2D
system is 
\begin{eqnarray}
\left( u_{t}+uu_{x}+u_{xxx}-v_{x}+\alpha u_{xx}+\gamma u_{xxxx}-\gamma
u_{yy}\right) _{x} &=&\pm u_{yy},  \label{uCDKP-II} \\
v_{t}+cv_{x}-u_{x}-\Gamma v_{xx} &=&0.  \label{vCDKP-II}
\end{eqnarray}
In fact, exactly the same combination of dissipative terms as in Eq. (\ref
{uCDKP-II}) has been earlier derived in asymptotic equations governing
nonlinear waves on thin downflowing liquid films \cite{Kawahara78,Melkonian}
and in two-fluid flows \cite{Goz}.\ 

Note that Eqs. (\ref{uCDKP-II}) and (\ref{vCDKP-II}) does not contain any
additional free parameter in comparison with the 1D system (\ref{u}), (\ref
{v}). As a matter of fact, this is another consequence of the fact that both
the dispersive and dissipative parts of the system were derived for the
generic quasi-1D case.

A special form of a 2D quasi-isotropic (rather than quasi-1D) generalization
of the Benney equation was also derived, which, for instance, describes
Rossby waves in a rotating atmosphere \cite{Rossby} (see also Refs. \cite
{Kawahara78,Melkonian}): 
\begin{equation}
u_{t}+uu_{x}+\Delta u_{x}+\alpha u_{xx}+\gamma \Delta ^{2}u=0,
\label{Benney}
\end{equation}
where $\Delta \equiv \partial _{x}^{2}+\partial _{y}^{2}$, hence $\Delta
^{2} $ in Eq. (\ref{Benney}) is a fourth-order isotropic dissipative
operator. Following the pattern of Eq. (\ref{Benney}), a quasi-isotropic
generalization of Eqs. (\ref{u}) and (\ref{v}) may be introduced, replacing
the term $u_{xxxx}$ in Eq. (\ref{usimplest}) by $\Delta ^{2}$, which leads
to a system 
\begin{eqnarray}
\left( u_{t}+uu_{x}+u_{xxx}-v_{x}+\alpha u_{xx}+\gamma \Delta ^{2}u\right)
_{x} &=&\pm u_{yy},  \label{uCDKP-I} \\
v_{t}+cv_{x}-u_{x}-\Gamma v_{xx} &=&0.  \label{vCDKP-I}
\end{eqnarray}
Note that all the 2D systems introduced above conserve two masses, 
\begin{equation}
M=\int_{-\infty }^{+\infty }\int_{-\infty }^{+\infty
}u(x,y)dx\,dy,\,N=\int_{-\infty }^{+\infty }\int_{-\infty }^{+\infty
}v(x,y)dx\,dy  \label{mass}
\end{equation}
(which indeed have the meaning of masses in the application to liquid-film
flows, in the cases of both the single layer with surfactant and two layers
with the lower one dominated by viscosity).

Thus, Eqs. (\ref{uCDKP-II}) and (\ref{vCDKP-II}) can be derived in a
consistent way as a system describing the downflow of a liquid viscous film
carrying a surfactant. Actually, the derivation outlined above clearly
suggests that this model is a generic one for weakly 2D systems combining
dispersion, gain, and viscosity. In particular, a derivation involving more
technicalities and following the lines of Ref. \cite{RMP} shows that the
same model applies to a downflow of a two-layered liquid film in the case
when the substrate layer is dominated by viscosity. A physically different
example may be a double-front flame propagating in a combustible gaseous
mixture, in the well-studied case when the combustion involves two
consecutive reactions (see Refs. \cite{double_flame} and references
therein). In the case when a single-flame front is unstable, it is well
known that its evolution is governed by the KS equation proper \cite{Grisha}.
It is known too that a situation with one front stable and one unstable is
possible, which is described by a linearly coupled system, consisting of a
KS equation and the one tantamount to Eq. (\ref{uCDKP-I}). Dispersion, which
is missing in the KS equation proper, can be induced by a background shear
flow tangent to the flame \cite{shear}, but detailed derivation of the full
model for this case is beyond the scope of this work.

As concerns the model (\ref{uCDKP-I}), (\ref{vCDKP-I}), we consider it as a
semi-phenomenological one, which may apply to cases which are ``more
isotropic'' than those obeying the condition (\ref{order}). We will study
this model parallel to the physical one (\ref{uCDKP-II}), (\ref{vCDKP-II})
in order to see if qualitative results are sensitive to details of a given
model. Accordingly, the systems (\ref{uCDKP-II}), (\ref{vCDKP-II}) and (\ref
{uCDKP-I}), (\ref{vCDKP-I}) will be below referred to as physical and
phenomenological ones, respectively. In the analysis presented in the
following sections, we focus on the case of the KP-I type, i.e., with the
upper sign in Eqs. (\ref{usimplest}), (\ref{uCDKP-II}) and (\ref{uCDKP-I}),
as only in this case one may expect the existence of nontrivial 2D pulses,
while the models of the KP-II type may only extend the SP found in Ref. \cite
{Malomed01} into a quasi-1D ($y$--independent) pulse in 2D.

\section{The stability of the zero solution}

As previously mentioned, completely stable SPs can only exist in a system
whose trivial solution, $u=v=0$, is stable, therefore our first objective is
to analyze this necessary stability condition. We substitute into the
corresponding linearized equations a 2D perturbation in the form $u\sim \exp
\left( ikx+iqy+\lambda t\right) $, $\,v\sim \exp \left( ikx+iqy+\lambda
t\right) $, where $k$ and $q$ are arbitrary real wave numbers of the
perturbation, and $\lambda $ is the corresponding instability growth rate
[note that, as all the equations that we are going to consider are written
in the scaled form, $k$ and $q$ are not assumed to be specially small,
unlike $K$ and $Q$ in Eq. (\ref{order})]. This leads to a linearized
dispersion equation 
\begin{equation}
\left[ k\left( \lambda -ik^{3}-\alpha k^{2}+\gamma k^{4}\right) -iq^{2}
\right] \left( \lambda +ick+\Gamma k^{2}\right) +k^{3}=0\,  \label{lambda1}
\end{equation}
for the formal model (\ref{usimplest}), (\ref{vsimplest}), or 
\begin{equation}
\left[ k\left( \lambda -ik^{3}-\alpha k^{2}+\gamma (k^{2}+q^{2})^{2}\right)
-iq^{2}\right] \left( \lambda +ick+\Gamma k^{2}\right) +k^{3}=0\,
\label{lambda2}
\end{equation}
for the phenomenological model (\ref{uCDKP-I}), (\ref{vCDKP-I}), or 
\begin{equation}
\left[ k\left( \lambda -ik^{3}-\alpha k^{2}+\gamma (k^{4}+q^{2})\right)
-iq^{2}\right] \left( \lambda +ick+\Gamma k^{2}\right) +k^{3}=0\,
\label{lambda3}
\end{equation}
for the physical model (\ref{uCDKP-II}), (\ref{vCDKP-II}). The stability
condition states that both solutions of the quadratic equations (\ref
{lambda1}), (\ref{lambda2}), or (\ref{lambda3}) must satisfy the inequality 
\begin{equation}
\mathrm{Re\,}\left[ \lambda (k,q)\right] \leq 0  \label{condition1}
\end{equation}
at all the real values of $k$ and $q$.

As it was already mentioned, the zero solution in the formal model (\ref
{usimplest}), (\ref{vsimplest}) is always unstable, which can be shown as
follows: in the case when $k$ is small, while $q$ is $\sim O(1)$, then two
roots of Eq. (\ref{lambda1}) can be expanded as 
\begin{equation}
\lambda _{1}(k)=-ick-\Gamma k^{2}+...\,,\,\,\lambda _{2}=iq^{2}/k+\alpha
k^{2}+\cdots .  \label{opposite}
\end{equation}
Obviously, the second root in Eq. (\ref{opposite}) yields instability.

The zero solution may be stable in the physical and phenomenological models.
Although direct check of the condition (\ref{condition1}) for Eqs. (\ref
{lambda2}) and (\ref{lambda3}) at all real values of $k$ and $q$ with the
four free parameters is a formidable algebraic problem, it is possible to
link the stability condition for the physical system to that which was
studied in detail for the 1D system (\ref{u}), (\ref{v}) in Ref. \cite
{Malomed01}. An algebraic transformation shows that, if the condition 
$\mathrm{Re\,}\left[ \lambda (k)\right] \leq 0$ holds at all real values of 
$k $ in Eq. (8) of Ref. \cite{Malomed01}, then the inequality $\mathrm{Re\,}
\left[ \lambda (k,q)\right] \leq 0$ is true at all real values of $k$ and $q$
in Eq. (\ref{lambda3}), or, in other words, the stability of the zero
solution in the 1D case guarantees its stability in the 2D case for the
physical model. However, rather than following formal algebra, it is easier
to understand this result from Fig. 1, which shows a 3D plot of the
instability growth rate $\mathrm{Re}\,\lambda $ vs. $k$ and $q$ for a set of
typical values of the parameters. Obviously, when $k$ is large, the
stability is secured by the higher-order dissipative term in Eq. (\ref
{uCDKP-II}). The most dangerous case is that when $k$ and $q$ are relatively
small. It can be seen from Fig. 1 that the growth rate, considered as a
function of $q$, attains its maximum value at $q=0$ if $k$ is very small. As 
$k$ increases, two additional local maxima of the growth rate appear at
nonzero $q$, and they may be greater than the local maximum at $q=0$.
However, since the local maximum at $q=0$ becomes more negative as $k$
increases, the two side maxima remain negative too. Thus, the zero solution
to Eqs. (\ref{uCDKP-II}) and (\ref{vCDKP-II}) is stable in the same
parameter region in which it was found to be stable in the 1D system 
(\ref{u}), (\ref{v}) in Ref. \cite{Malomed01}.

In the phenomenological model, the zero solution is also stable in a certain
parametric region. However, no simple relation of the stability condition to
that in the 1D system can be found in this model.

\section{The perturbation theory for two-dimensional solitary pulses}

Both the physical and phenomenological models reduce to a zero-order system
by setting $\alpha =\gamma =\Gamma =0$, while keeping an arbitrary value of 
$c$. This zero-order system is conservative, consisting of the KP-I equation
coupled to an extra linear one, 
\begin{equation}
\left( u_{t}+uu_{x}+u_{xxx}-v_{x}\right) _{x}=u_{yy}\,,\,v_{t}+cv_{x}=u_{x}.
\label{zero_order}
\end{equation}

Looking for a solution to Eqs. (\ref{zero_order}) in the form of a soliton
traveling at a constant velocity $s$ in the $x$-direction, so that 
\begin{equation}
u(x,y,t)=u(\xi ,y),\,v(x,y,t)=v(\xi ,y),\,\mathrm{with\,\,\,}\xi \equiv x-st,
\label{xi}
\end{equation}
we immediately conclude that, as well as in the 1D case, for such a solution
we have 
\begin{equation}
v(\xi ,y)=\left( c-s\right) ^{-1}u(\xi ,y).  \label{uv}
\end{equation}
With regard to the relation (\ref{uv}), an \emph{exact} solution to Eqs. 
(\ref{zero_order}) giving a 2D weakly localized soliton (``lump'') is 
\begin{equation}
u(\xi ,y)=\frac{24\left( 1+cs-s^{2}\right) }{c-s}\cdot w(\zeta ,z)\,,
\label{lump}
\end{equation}
where 
\begin{equation}
z\equiv \frac{1+cs-s^{2}}{c-s}\cdot y\,,\,\,\,\zeta \equiv \sqrt{\frac{
1+cs-s^{2}}{c-s}}\cdot \xi \,,\,\,\,w(\zeta ,z)\equiv \frac{z^{2}-\zeta
^{2}+3}{\left[ z^{2}+\zeta ^{2}+3\right] ^{2}}\,.  \label{zzeta}
\end{equation}
This solution exists in the case 
\begin{equation}
\frac{1+cs-s^{2}}{c-s}>0\,.  \label{condition}
\end{equation}

Note that, in contrast to the lump soliton solutions of the KP-I equation
proper, which may only have positive velocities, the solitons (\ref{lump})
may also move in the negative direction, provided that the velocity
satisfies the condition (\ref{condition}). In the particular case $c=0$,
this condition means that either $s>1$, or $-1<s<0$.

We have checked by direct simulations of Eqs. (\ref{zero_order}) that all
the lump-type SP solutions are stable within the framework of the
unperturbed equations (\ref{zero_order}). We have also simulated catch-up
and head-on collisions between two lump pulses with different velocities.
The results clearly show that the collisions are \emph{inelastic}, hence, on
the contrary to the KP-I equation, the conservative system (\ref{zero_order})
is not an exactly integrable one. An example of the head-on collision
between two lump-type SPs with velocities $s=2.0$ and $s=-0.8$ is shown in
Fig. 2. The inelastic character of the interaction is obvious.

In fact, the collision shown in Fig. 2 is not quite generic, as it is
generated by an initial configuration in which the mass $N$ of the 
$v$--component, see Eq. (\ref{mass}), is almost zero ($N$ vanishes exactly if
the initial velocity of the second soliton is$\allowbreak -0.756$ instead of 
$-0.8$, which is the case in Fig. 2). In this case, a resonance probably
happens, resulting in generation of an additional pair of solitons, which is
suggested by Fig. 2 [especially, by the panel 2(d)]. Far from this resonant
case, a typical collision (see an example in Fig. 3) is also strongly
inelastic, but in this case the result may be regarded, in the first
approximation, as fusion of two solitons into one.

To check the accuracy of the results produced by direct simulations of the
2D equations, including those displayed in Figs. 2 and 3 and in the next
section, the simulations were repeated increasing the number of points. For
instance, Figs. 2 and 3 were obtained in a 2D domain of $256\times 256$
points; repeating the same simulations with $512\times 512$ points, we had
obtained the same pictures, without any visible difference.

Next, we restore the terms with the coefficients $\alpha $, $\gamma $ and 
$\Gamma $ as small perturbations, with the aim to predict, at the first order
of the perturbation theory, which solutions will be selected from the family
of the lump solitons (\ref{lump}) found in the zero-order system (\ref
{zero_order}). As it is suggested by the analysis of the 1D system \cite
{Malomed01}, it is convenient to choose the relative velocity, $\delta
\equiv c-s$, as a parameter of the family.

As well as in the 1D case, we select stationary pulses, using the balance
condition for the net field momentum, 
\begin{equation}
P=\frac{1}{2}\int_{-\infty }^{+\infty }dx\int_{-\infty }^{+\infty }dy\,\left[
u^{2}(x,y)+v^{2}(x,y)\right] ,  \label{P}
\end{equation}
which is a dynamical invariant of the conservative (zero-order) system [the
masses (\ref{mass}) cannot be used for this purpose, as they remain
dynamical invariants in the full dissipative models]. The dissipative and
gain terms give rise to the following exact balance equations for the
momentum: 
\begin{equation}
\frac{dP}{dt}=\int_{-\infty }^{+\infty }dx\int_{-\infty }^{+\infty }dy\,
\left[ \alpha u_{x}^{2}-\gamma (u_{xx}^{2}+u_{y}^{2})-\Gamma v_{x}^{2}\right]
\,  \label{balance2}
\end{equation}
in the physical system (\ref{uCDKP-II}), (\ref{vCDKP-II}), and 
\begin{equation}
\frac{dP}{dt}=\int_{-\infty }^{+\infty }dx\int_{-\infty }^{+\infty }dy\,
\left[ \alpha u_{x}^{2}-\gamma (u_{xx}+u_{yy})^{2}-\Gamma v_{x}^{2}\right] \,
\label{balance1}
\end{equation}
in the phenomenological system (\ref{uCDKP-I}), (\ref{vCDKP-I}).

Substituting the unperturbed exact solution (\ref{lump}) into the
equilibrium condition, $dP/dt=0$, which follows from Eqs. (\ref{balance2})
or (\ref{balance1}), we cast the equilibrium condition in the form 
\begin{equation}
C_{1}\left( \alpha -\frac{\Gamma }{\delta ^{2}}\right) \left( c-\delta +
\frac{1}{\delta }\right) -C_{2}\left( c-\delta +\frac{1}{\delta }\right)
^{2}=0,  \label{balance_eq}
\end{equation}
where, for both the physical and phenomenological systems, 
\begin{equation}
C_{1}\equiv \int_{-\infty }^{+\infty }dz\int_{-\infty }^{+\infty }d\zeta
\left( \frac{\partial w}{\partial \zeta }\right) ^{2},  \label{C_1}
\end{equation}
and the constant $C_{2}$ is defined differently for the two models: 
\begin{equation}
C_{2}^{\mathrm{(phys)}}=\int_{-\infty }^{+\infty }dz\int_{-\infty }^{+\infty
}d\zeta \left[ \left( \frac{\partial ^{2}w}{\partial \zeta ^{2}}\right)
^{2}+\left( \frac{\partial w}{\partial z}\right) ^{2}\right] \,,
\label{C_2b}
\end{equation}
\begin{equation}
C_{2}^{\mathrm{(phen)}}=\int_{-\infty }^{+\infty }dz\int_{-\infty }^{+\infty
}d\zeta \left( \frac{\partial ^{2}w}{\partial \zeta ^{2}}+\frac{\partial
^{2}w}{\partial z^{2}}\right) ^{2}.  \label{C_2a}
\end{equation}
Numerical computation of the integrals yields $C_{1}=0.34708$, $C_{2}^{
\mathrm{(phys)}}=0.75984$, and $C_{2}^{\mathrm{(phen)}}=0.85770$. Then, the
equilibrium equation (\ref{balance_eq}) yields the following cubic equation
for $\delta $ (where $\widetilde{\alpha }\equiv \alpha /\gamma $, 
$\widetilde{\Gamma }\equiv \Gamma /\gamma $): 
\begin{equation}
\delta ^{3}+\left( 0.46\widetilde{\alpha }-c\right) \delta ^{2}-\delta -0.46
\widetilde{\Gamma }=0,  \label{cubic2}
\end{equation}
\begin{equation}
\delta ^{3}+\left( 0.41\widetilde{\alpha }-c\right) \delta ^{2}-\delta -0.41
\widetilde{\Gamma }=0;  \label{cubic1}
\end{equation}
for the physical and phenomenological models, respectively. Note that these
equations have the same general form as the one which selects equilibrium
values of the parameter $\delta $ in the 1D system investigated in Ref. \cite
{Malomed01}.

Physical roots of Eq. (\ref{cubic2}) or (\ref{cubic1}) are defined as those
which not only are real, but also satisfy the condition (\ref{condition}).
Physical roots select particular lump-type SPs, from the family of lump
solitons of the zero-order system, that remain steady pulses in the presence
of the perturbations. As well as in the 1D model, the existence of at least
two physical roots is necessary for one pulse to be stable (the other one,
with a smaller value of the amplitude, is then automatically unstable) \cite
{Malomed01}. Numerical solution of Eqs. (\ref{cubic2}) and (\ref{cubic1})
verifies that both equations have exactly two physical roots in a broad
parametric region, while three physical roots can never be found.
Interestingly, the velocities corresponding to these two physical roots are
both positive, i.e., a stable lump SP cannot move in the negative 
$x$--direction.

Regions in the parameter plane ($\alpha ,\Gamma $) in which exactly two
physical roots have been found are shown, for $\gamma =0.05$, in Figs. 4 and
5 for the physical model, with $c=0$ and $c=-1$, respectively, and in Fig. 6
for the phenomenological model with $c=-1$. These regions are bounded by two
solid lines in Figs. 4 through 6. Beneath the lower solid line, Eq. (\ref
{cubic2}) or (\ref{cubic1}) has three real roots, but no more than one of
them is physical, and above the upper solid line, two physical roots
bifurcate into a pair of complex ones, leaving no physical roots.

The dashed line is the border above which the zero solution is stable in the
1D model (\ref{u}), (\ref{v}), hence it is also stable in the 2D physical
model, according to the results reported in the previous section. As is seen
in Fig. 4, in the parameter plane ($\alpha ,\Gamma $) of the physical system
with $c=0$, there is no overlapping between regions where the zero solution
is stable, and where the perturbation theory selects two physical solutions
for the 2D pulses, but a narrow overlapping region is found in the same
model for $c=-1$, see Fig. 5. It is expected that stable lump-type SPs exist
inside this narrow region, which is confirmed by direct simulations, as
reported in the next section. Despite the existence of an overlapping region
in Fig. 6, we have to check the condition (\ref{condition}) for Eq. (\ref
{lambda2}), i.e., the stability of the zero solution in the phenomenological
model, point by point, since the stability of the zero solution in the 1D
model does not guarantee the stability of the zero solution in the
phenomenological system, see the previous section.

The same analysis can be performed for values of the loss parameter $\gamma $
other than $0.05$, which was fixed in the above consideration. The results
show that the variation of $\gamma $ produces little change in terms of the
expected SP stability region. On the other hand, the group-velocity mismatch 
$c$ affects the stability region significantly. We had numerically found
that there is no stability region in both physical and phenomenological
model for $c\geq 0$. Then, by inspecting the region of negative $c$, it was
found that there is a critical value $c_{\mathrm{cr}}$ such that stable
pulses become possible for $c<c_{\mathrm{cr}}$. In the physical model, 
$c_{\mathrm{cr}}\approx -0.8$.

To conclude the analytical consideration, it is necessary to stress that the
stability conditions obtained here are only necessary ones. Obtaining a full
set of sufficient stability conditions might be possible within the
framework of a rigorous spectral analysis of the full system linearized
around the pulse solution, which was done (in a numerical form) 
for the 1D Benney (KS -- KdV)
equation in Refs. \cite{Chang}. Extending this analysis to the two-component
2D case is an extremely difficult problem, which is beyond the scope of the
present work. Nevertheless, numerical results reported in the next section
strongly suggest that the necessary stability conditions, that were obtained
above by means of the simple analytical methods, are, most probably,
sufficient for the stability of the 2D pulses.

\section{Direct simulations of the two-dimensional solitary pulses}

To check predictions produced by the above analysis for both the physical
and phenomenological models, against direct numerical simulations, we have
upgraded the implicit pseudo-spectral scheme, which was developed previously
for the 1D model (\ref{u}), (\ref{v}) (see Appendix in Ref. \cite{Malomed01}),
to a scheme capable of dealing with both 2D models. It it relevant to
mention that results obtained with a rather coarse grid, $\Delta x=\Delta
y=0.5$, and relatively large time steps, $\Delta t=0.1$, remained virtually
unchanged if reproduced with an essentially finer grid and smaller time
step. The initial conditions were taken as the lump-soliton solutions of the
zero-order system, but with arbitrarily modified values of the amplitude, in
order to check whether strongly perturbed pulses relax to stable ones, i.e.,
whether the stable pulses are \textit{attractors}.

Collecting data produced by the systematic simulations, it has been found
that, for the physical model, stable lump-type SPs do exist and are stable 
\emph{everywhere} inside the stability region in the ($\alpha ,\Gamma $)
parameter plane which was predicted by the analytical perturbation theory,
see Fig. 5. Moreover, all the stable pulses were found to be strong
attractors indeed. A typical lump-type SP with $\alpha =0.2$, $\gamma =0.05$,
$c=-1.0$, and $\Gamma =0.55$ is displayed in Fig. 7. The initial-state
pulses selected as mentioned above definitely relax to this single
stationary lump-type SP, provided that the initial amplitude $A_{0}$ of
their $u$-component exceeds $1.5$. For instance, starting with the initial
amplitudes $A_{0}=1.96$ and $A_{0}=13.33$ [the corresponding values of the
unperturbed-solitons' velocity are $s_{0}=0.8$ and $s_{0}=2.0$, see Eq. (\ref
{lump})], a lump-type SP develops with the values of the amplitude $A_{
\mathrm{num}}=7.11$ and $A_{\mathrm{num}}=7.16$, and velocities $s_{\mathrm{
\ num}}=1.32$ and $s_{\mathrm{num}}=1.33$, respectively, by the moment of
time $t=400$. Meanwhile, the analytical prediction for the amplitude and
velocity of the presumably stable pulse, given by Eq. (\ref{cubic2}) for the
same values of the parameters, is $A_{\mathrm{anal}}\approx 7.38$ and $s_{
\mathrm{anal}}\approx 1.35$. Thus, the numerical results match the
theoretical prediction well.

On the other hand, if the initial amplitude is too small, e.g., $A_{0}=0.89$
($s_{0}=0.7$), the pulse decays to zero, which is natural too, as the stable
zero solution has its own attraction basin. Note that for the second
(smaller) steady-state pulse, which is expected to play the role of the
separatrix between the attraction basins of the stable pulse and zero
solution, the perturbation theory predicts, in the same case, the amplitude 
$\widetilde{A}_{\mathrm{anal}}\approx 1.26$, so it seems quite natural that
the initial pulses with $A_{0}=1.96$ and $A_{0}=0.89$ relax, respectively,
to the stable pulse and to zero.

For the phenomenological model, no definite stability region for the pulses
has been found above; instead, the stability of the zero solution at each
point inside the region bounded by the solid lines in Fig. 6 should be
checked separately. For instance, the zero solution is found to be stable at
the parameter point $(\alpha =0.2,\Gamma =0.55)$ in Fig. 6. At this point,
the two physical roots of Eq. (\ref{cubic1}) are $2.073$ and $1.786$. The
amplitudes of the corresponding solitons predicted by the perturbation
theory are $\widetilde{A}_{\mathrm{anal}}\approx 4.72$ and $\widetilde{A}_{
\mathrm{anal}}\approx 1.81$, respectively; the one with the larger amplitude
may be stable according to the general principle formulated above.
Simulations demonstrate that, indeed, there is a stable 2D pulse with a
shape very close to the predicted one, all the initial states in the form of
the conservative-model lump, whose amplitude exceeds $2.0$, relaxing to this
stable SP. Figure 8 shows the shapes of the $u$-- and $v$--components of the
SP at $t=400$, generated by the initial configuration taken as the soliton
of the zero-order system (\ref{zero_order}) with the amplitude $13.33$. The
amplitude of the thus obtained $u$-components is about $5.14$, the
discrepancy with the analytical prediction being less than $9\%$.

It was also found that all the initial lump pulses whose amplitude is less
than $1.6$ decay to zero. This fact complies well with the expectation that
the second (smaller-amplitude) steady-state pulse, predicted by the
perturbation theory, whose amplitude is $1.81$, ought to be the separatrix
between the attraction basins of the stable pulse and zero solution in the
phenomenological system.

The simulations also show that, even if the zero background is unstable,
lump-type SPs may be quite stable when the integration domain (supplemented
by periodic boundary conditions in both $x$ and $y$) is not very large. An
explanation given in Ref. \cite{Malomed01} to a similar ``overstability''
effect found in the 1D model, which is based on suppression of nascent
perturbations by the pulse periodically traveling around the domain, applies
to the 2D case as well.

Numerical simulations of interactions between two or more stable lump-type
SPs were also carried out. Unlike the 1D system (\ref{u}), (\ref{v}), in
which stable bound states of two and three pulses were easily found \cite
{Malomed01} (and unlike a physically relevant 2D model of a different type,
which follows the pattern of the Zakharov-Kuznetsov equation \cite{ZK} and
will be considered elsewhere), bound states of pulses have \emph{not} been
found in both the physical and phenomenological systems based on the KP-I
equation. Actually, this fact may be benign for experimental observation of
the 2D pulses, as there will be no competition with multi-humped structures
that might render the picture much more complex.

\section{Conclusion}

In this work, we have extended the 1D stabilized Kuramoto - Sivashinsky
system to the 2D case. The 2D model is, quite naturally, based on the
Kadomtsev-Petviashvili (KP) equation supplemented by loss and gain terms. A
model with the dissipative terms of the Newell-Whitehead-Segel type was
derived in a consistent way for the weakly 2D case. The derivation was
outlined for a particular physical system, a downflowing liquid film
carrying a surfactant that diffuses along its surface. It was argued that,
in fact, the derived model is generic, therefore it also applies, for
instance, to the description of interfacial waves in a two-layered flowing
film, and double-front flames in a gaseous mixture. Another model, which may
apply to a more isotropic situation, was put forward as a
semi-phenomenological one. Both models consist of the generalized KP-I
equation with gain and loss terms, linearly coupled to an extra equation of
the advection-diffusion type. The additional linear equation stabilizes the
system's zero solution, thus paving the way to the existence of completely
stable 2D localized solitary pulses, which are objects of an obvious
physical interest. Treating the losses and gain as small perturbations, and
employing the balance equation for the net field momentum, we have found
that the condition of the equilibrium between the losses and gain may select
two steady-state 2D solitons from their continuous family, which was found,
in an exact analytical form, in the absence of the loss and gain (the exact
solitons are similar to the ``lump'' solutions of the KP-I equation). When
the zero solution is stable and, simultaneously, two lump-type pulses are
picked up by the balance equation for the momentum, the pulse with the
larger value of the amplitude is expected to be stable in the infinitely
long system, while the other pulse must be unstable, playing the role of a
separatrix between attraction domains of the stable pulse and zero solution.
These predictions have been completely confirmed by direct simulations for
both the physical and phenomenological systems. Another noteworthy finding
is that, unlike their 1D counterpart, both KP-based 2D systems do not
generate stable bound states of pulses.

\newpage

\begin{center}
\textbf{Figure Captions}
\end{center}

Fig. 1. The instability growth rate $\mathrm{Re}\,\lambda $ for the zero
solution in the physical model [Eqs. (\ref{uCDKP-II}) and (\ref{vCDKP-II})]
vs. the longitudinal and transverse perturbation wave numbers $k$ and $q$.
The values of the parameters are $\alpha =0.2$, $\gamma =0.05$, $c=-1.0$,
and $\Gamma =0.55$.

Fig. 2. An example of the inelastic collision between two stable lump-type
solitons with different velocities, simulated within the framework of the
zero-order conservative system (\ref{zero_order}) with $c=0$. The $u$- and $v
$-fields prior to the collision (at $t=0$) are shown in panels (a) and (b),
and after the collision (at $t=40$) -- in panels (c) and (d). The initial
velocities of the two solitons are $s_{1}=2.0$ \ and $s_{2}=-0.8$.

Fig. 3. The inelastic collision in the conservative system (\ref{zero_order})
with $c=0$ in the case when the initial velocities of the two solitons are 
$s_{1}=2.0$ \ and $s_{2}=-0.5$. The panels have the same meaning as in Fig.
2, with a difference that the final state is displayed for $t=45$.

Fig. 4. The stability region of the zero solution, and the region inside
which two physical roots exist for Eq. (\ref{cubic2}) with $\gamma =0.05$
and $c=0$. The zero solution is stable above the dashed line, and inside the
region bounded by two solid lines, Eq. (\ref{cubic2}) produces two physical
solutions.

Fig. 5. The expected stability region for the solitary pulses in the
parametric plane ($\alpha $,$\Gamma $) for $\gamma =0.05$ and $c=-1.0$. The
solid and dashed lines have the same meaning as in Fig. 4.

Fig. 6. The stability region of the zero solution of the 1D model (\ref{u}),
(\ref{v}) and the region inside which two physical roots exist for Eq. (\ref
{cubic1}) when $\gamma =0.05$ and $c=-1$. The solid and dashed lines have
the same meaning as in Fig. 4.

Fig. 7. A stable lump-type solitary pulse in the physical model, found as a
result of long numerically simulated evolution in the case $\alpha =0.2$, 
$\gamma =0.05$, $c=-1.0$, and $\Gamma =0.55$. The panels (a) and (b) show
established shapes of the $u$ and $v$ fields, respectively.

Fig. 8. A stable lump-type solitary pulse in the phenomenological model,
found the same way as the pulse shown in Fig. 7 for the same values of the
parameters. The panels (a) and (b) have the same meaning as in Fig. 7.


\begin{thebibliography}{99}
\bibitem{Ablowitz}  M.~J.~Ablowitz and P.~A.~Clarkson, \emph{Solitons,
Nonlinear Evolution Equations and Inverse Scattering} (Cambridge University
Press, Cambridge (UK), 1991).

\bibitem{CH}  I. S. Aranson and L. Kramer, Rev. Mod. Phys. \textbf{74}, 99
(2002).

\bibitem{Benney}  D. J. Benney, J. Math. Phys. \textbf{45}, 150 (1966).

\bibitem{KSKdV}  T. Kawahara and S. Toh, Phys. Fluids \textbf{28}, 1636
(1985); Pure Appl. Math. \textbf{43}, 95 (1989); C. Elphick, G. R. Ierley,
O. Regev, and E. A. Spiegel, Phys. Rev. E \textbf{44}, 1110 (1991); A. Oron
and D. A. Edwards, Phys. Fluids A \textbf{5}, 506 (1993); C. I. Christov and
M. G. Velarde, Physica D \textbf{86}, 323 (1995).

\bibitem{RMP}  A. Oron, S. H. Davis, and S. G. Bankoff, Rev. Mod. Phys. 
\textbf{69,} 931 (1997).

\bibitem{SP}  S. Toh and T. Kawahara, J. Phys. Soc. Jpn. \textbf{54}, 1257
(1985); H.-C. Chang, Phys. Fluids \textbf{29}, 3142 (1986); T. Kawahara and
S. Toh, Phys. Fluids \textbf{31}, 2103 (1988); H.-C. Chang, E. A. Demekhin,
and D. I. Kopelevich, Phys. Rev. Lett. \textbf{75}, 1747 (1995).

\bibitem{Chang}  H.-C. Chang, E.A. Demekhin, and D.I. Kopelevich, Physica D 
\textbf{97}, 353 (1996); H.-C. Chang, E.A. Demekhin, and E. Kalaidin, SIAM\
J. Appl. Math. 58, 1246 (1998); H.-C. Chang and E.A. Demekhin, J. Fluid
Mech. \textbf{380}, 233 (1999).

\bibitem{Malomed01}  B.~A.~Malomed, B.-F.~Feng, and T.~Kawahara, Phys. Rev.
E \textbf{64}, 6304 (2001).

\bibitem{spot}  J. Schumacher and B. Erckhardt, Phys. Rev. E 63, 046307
(2001); A. Das and J. Mathew, Computers \& Fluids \textbf{30}, 533 (2001).

\bibitem{Petviashvili70}  B.~B. Kadomtsev and V.~I.~Petviashvili, Sov. Phys.
Dokl. \textbf{15}, 539 (1970).

\bibitem{Velarde94}  A. A. Nepomnyashchy and M. G.~ Velarde, Phys. Fluids 
\textbf{6}, 187 (1994); G. Huang, M. G. Velarde, and V. N. Kudryavtsev,
Phys. Rev. E \textbf{57}, 5473 (1998).

\bibitem{NWS}  L. A. Segel, J. Fluid Mech. \textbf{38}, 203 (1969); A. C.
Newell and J. A. Whitehead, \textit{ibid}. \textbf{38}, 279 (1969).

\bibitem{Kawahara78}  J.~Topper and T.~Kawahara, J. Phys. Soc. Jpn. \textbf{
\ 44}, 663 (1978); S.~Toh, H.~Iwasaki, T.~Kawahara, Phys. Rev. A \textbf{40},
5472 (1989).

\bibitem{Melkonian}  S.~Melkonian and S.~A.~Maslowe, Physica D \textbf{34},
255 (1989).

\bibitem{Goz}  M. F.~G\"{o}z, Physica D \textbf{123}, 112 (1998).

\bibitem{Rossby}  V. I.~Petviashvili, JETP Lett. \textbf{32} 619 (1980).

\bibitem{double_flame}  G. Joulin and P. Clavin, Combust. Flame \textbf{25},
389 (1975); L. Pelaez and A. Li\~{n}an, SIAM J. Appl. Math. \textbf{45}, 503
(1985); J. Pelaez, \textit{ibid}. \textbf{47}, 781 (1987).

\bibitem{Grisha}  G. I. Sivashinsky, Ann. Rev. Fluid Mech. \textbf{15}, 179
(1983).

\bibitem{shear}  Y. Kortsarts, I. Kliakhandler, L. Shtilman, and G. I.
Sivashinsky, Quart. Appl. Math. \textbf{56}, 401 (1998).

\bibitem{ZK}  V. E.~ Zakharov and E. A.~Kuznetsov, Sov. Phys. JETP \textbf{39},
285 (1974).
\end{thebibliography}
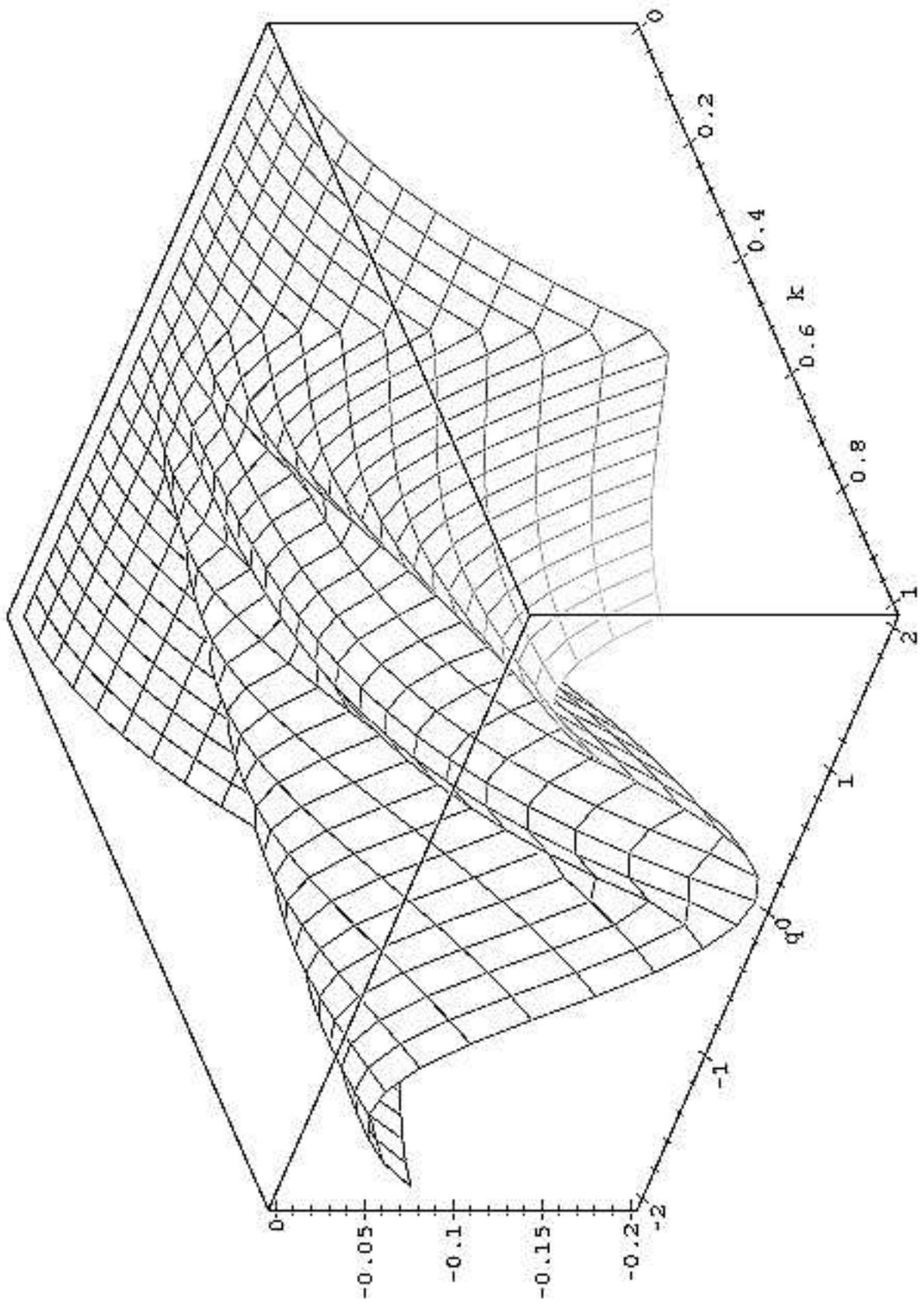
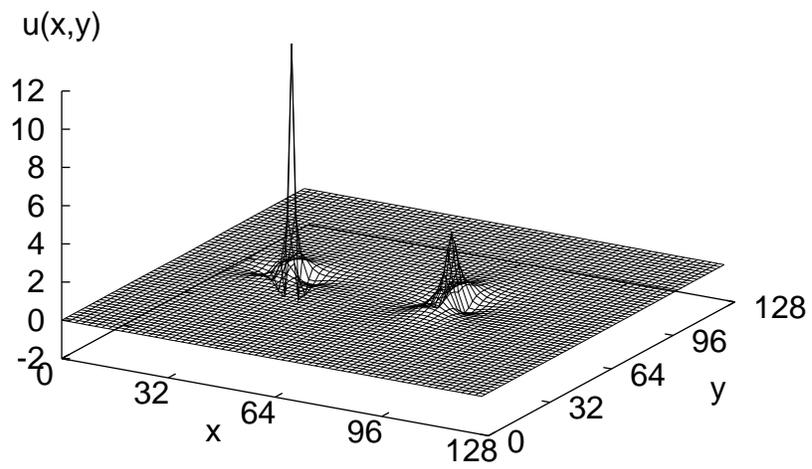
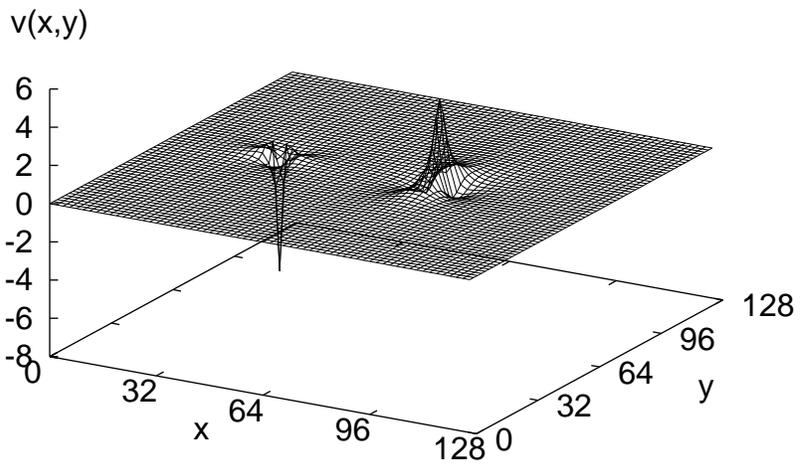
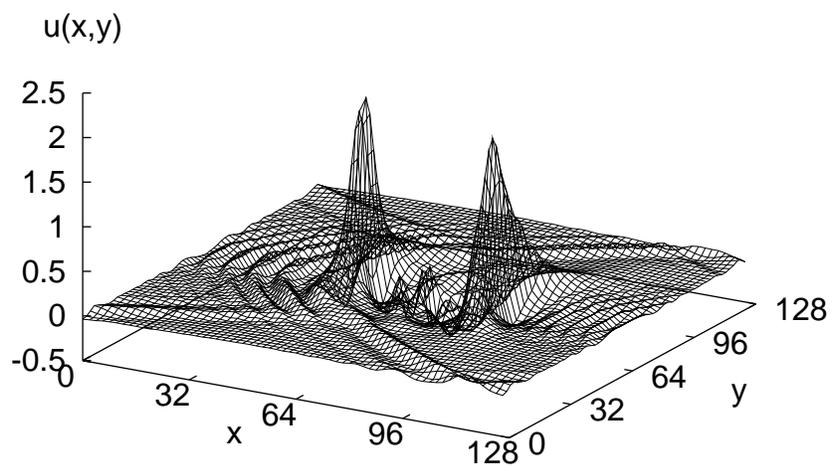
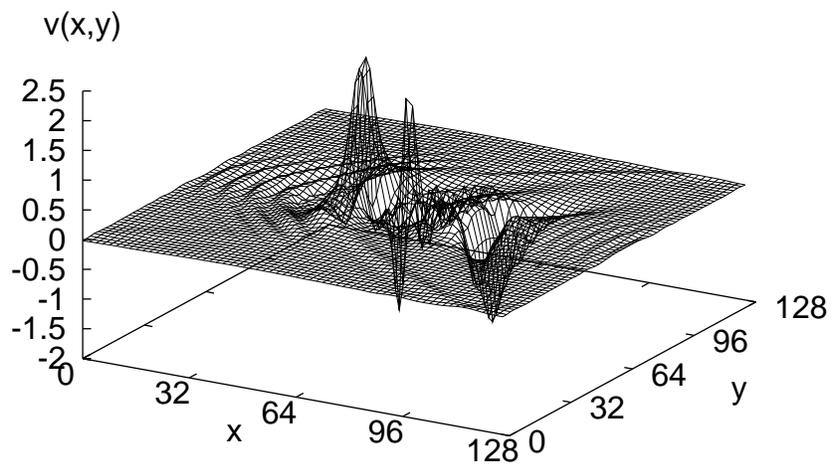
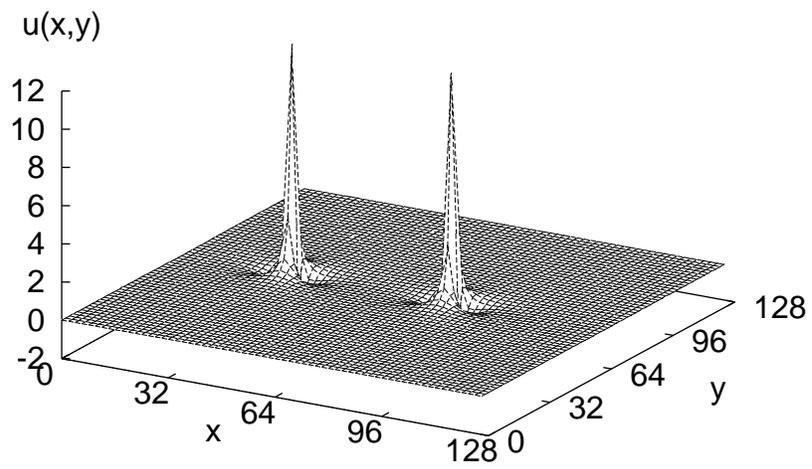
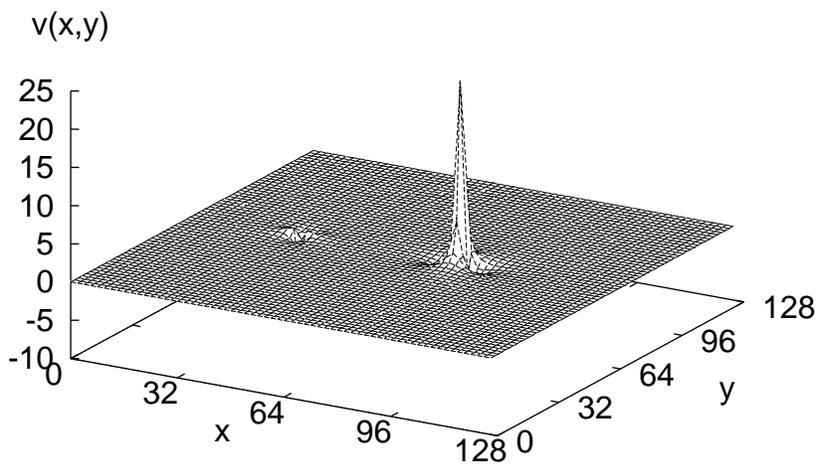
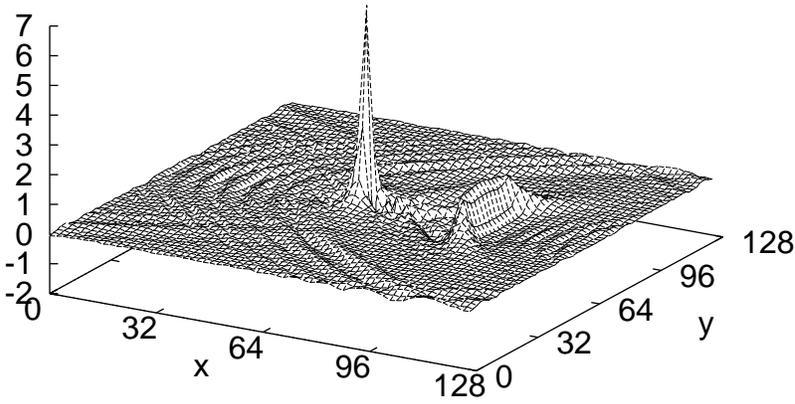
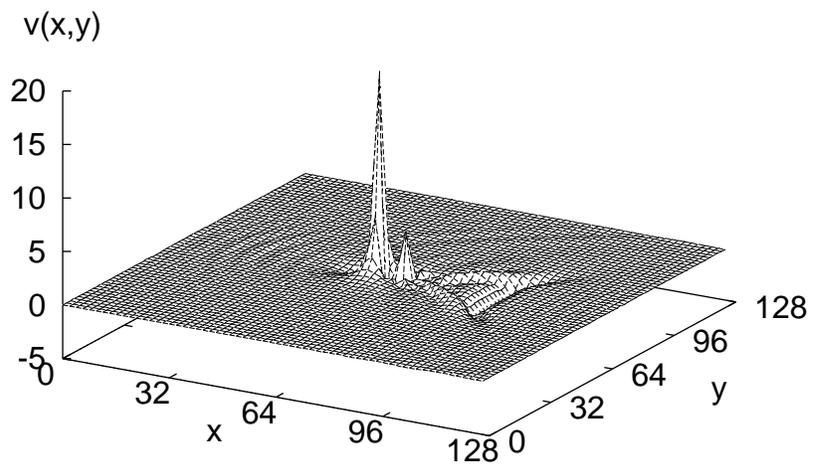
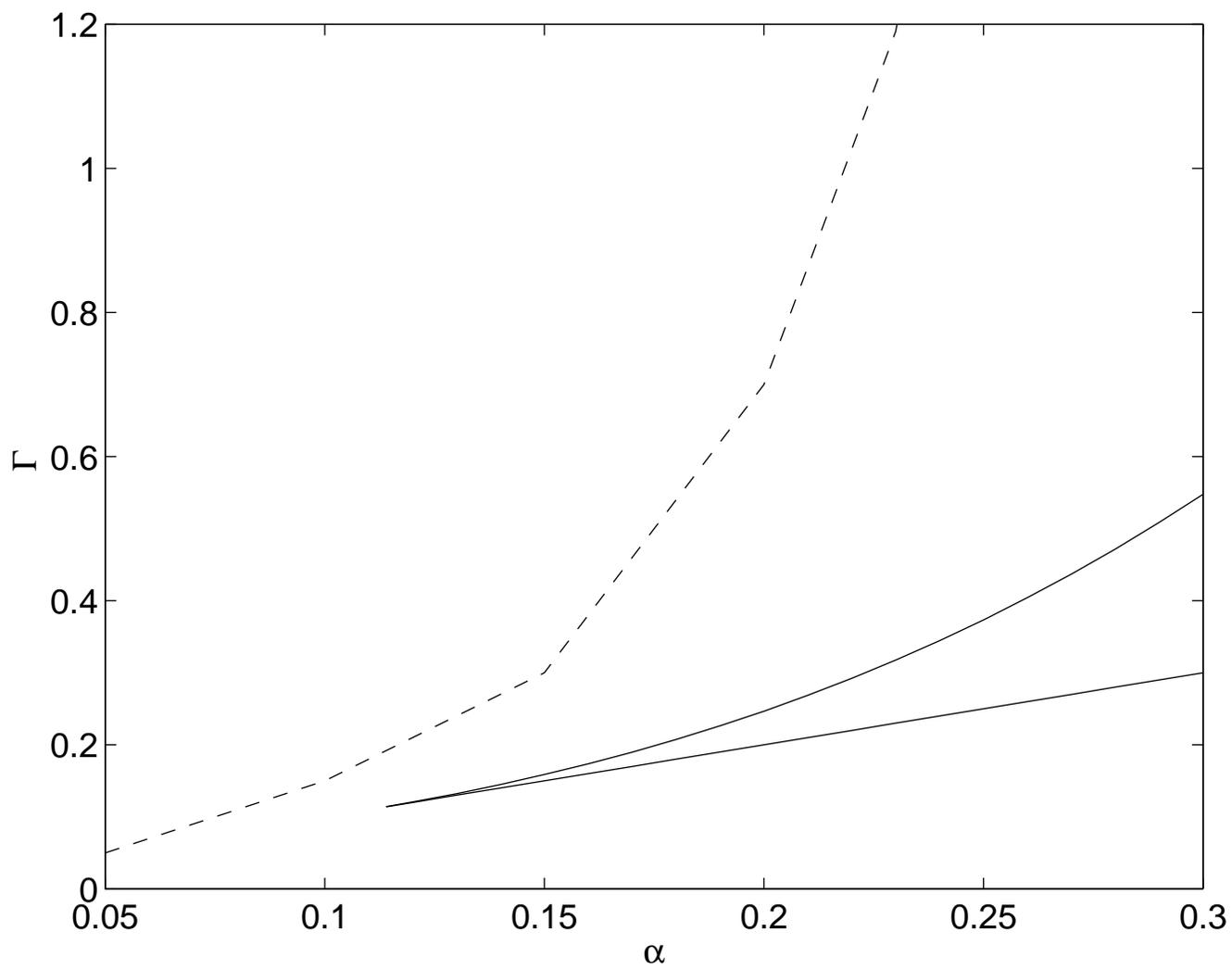
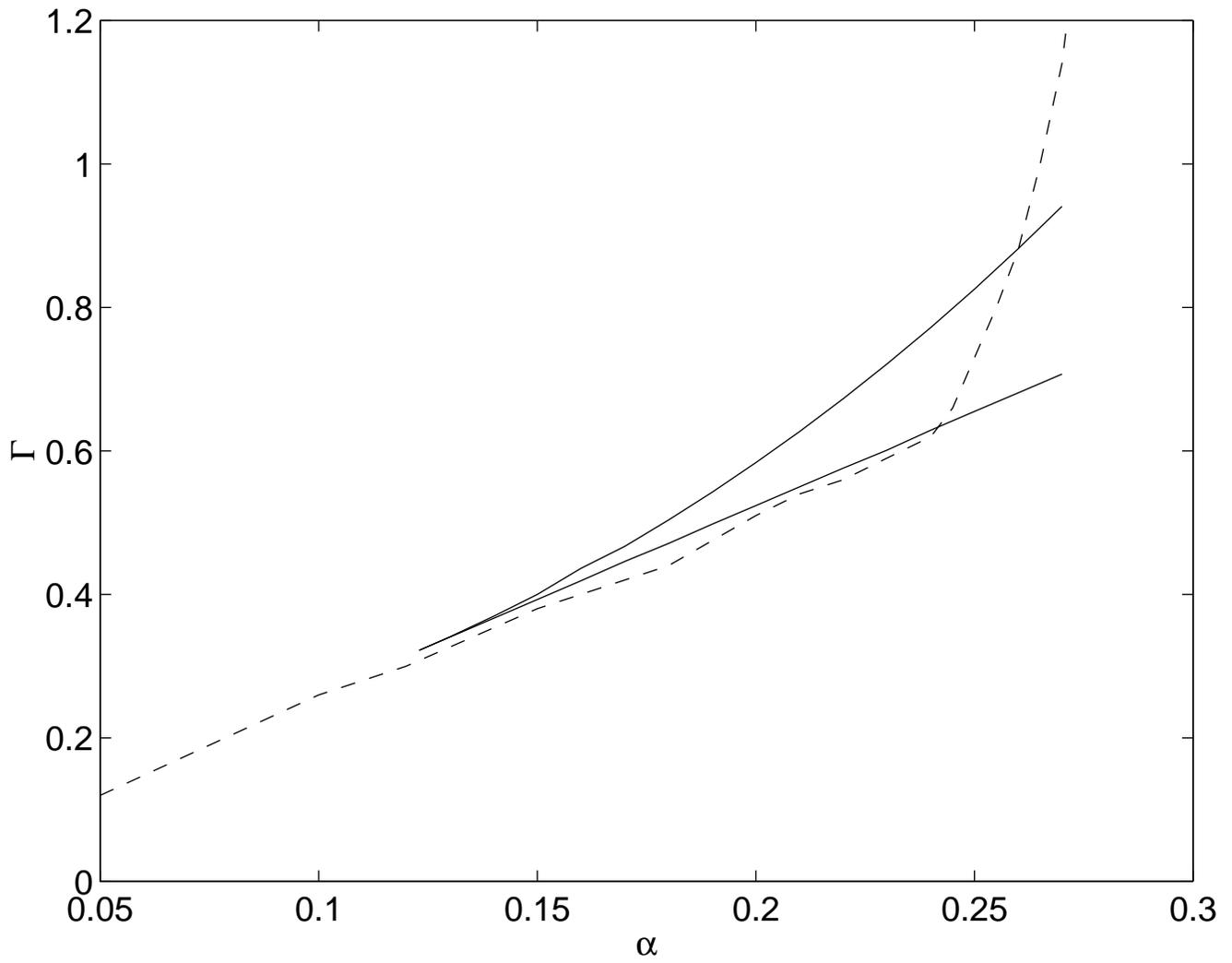
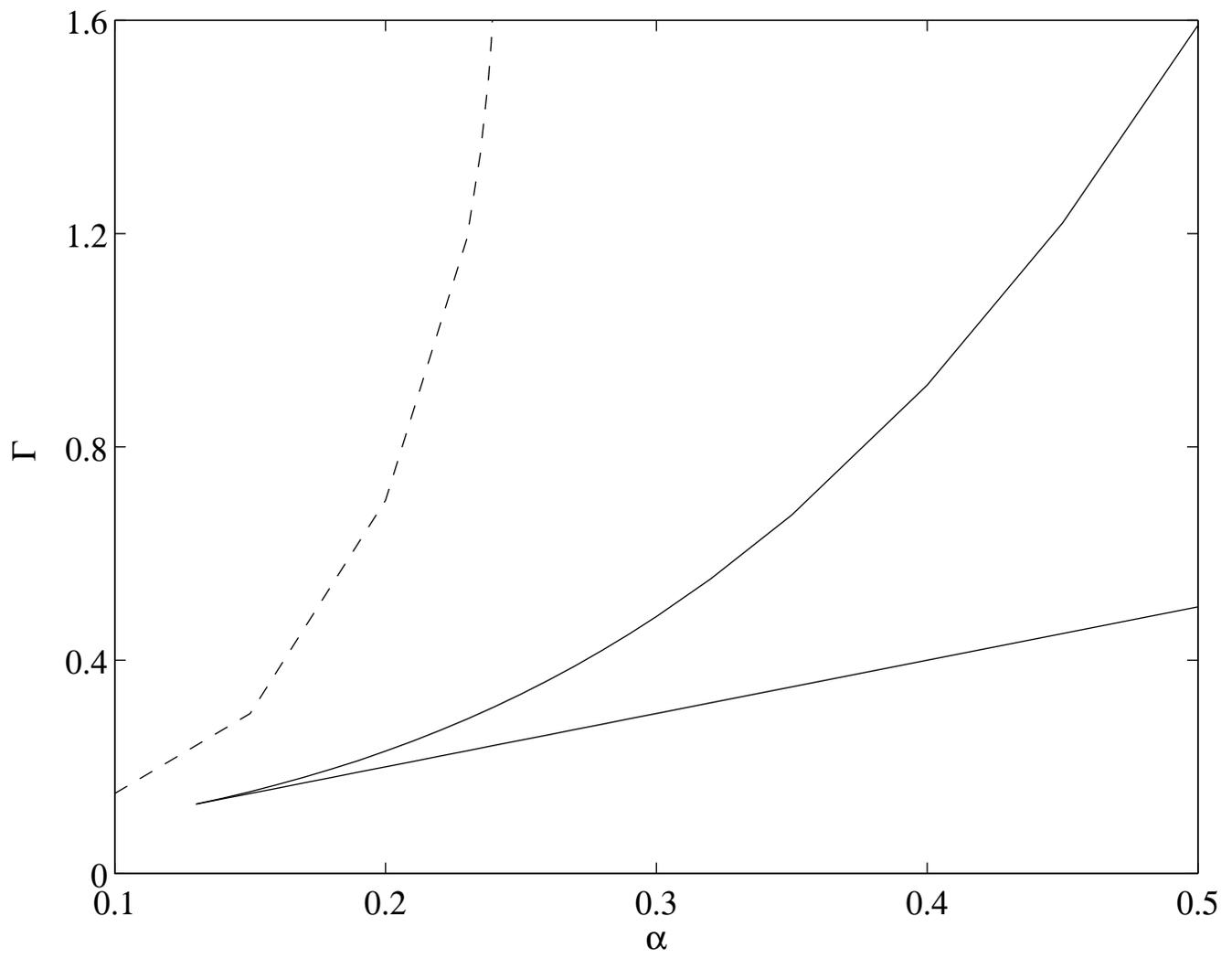
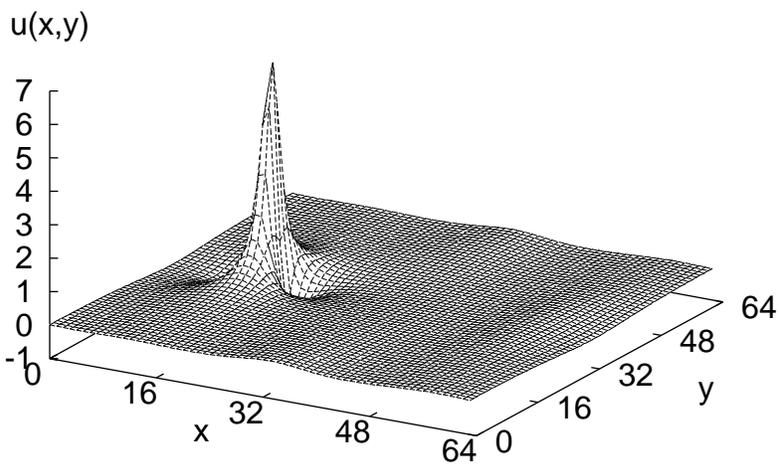
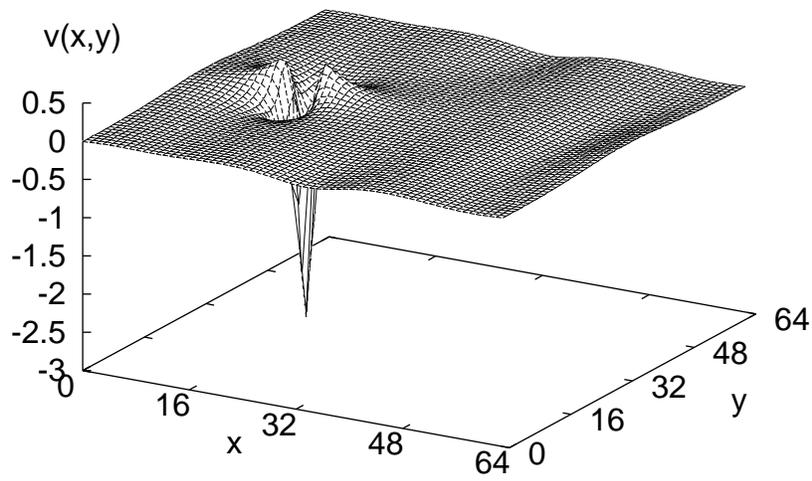
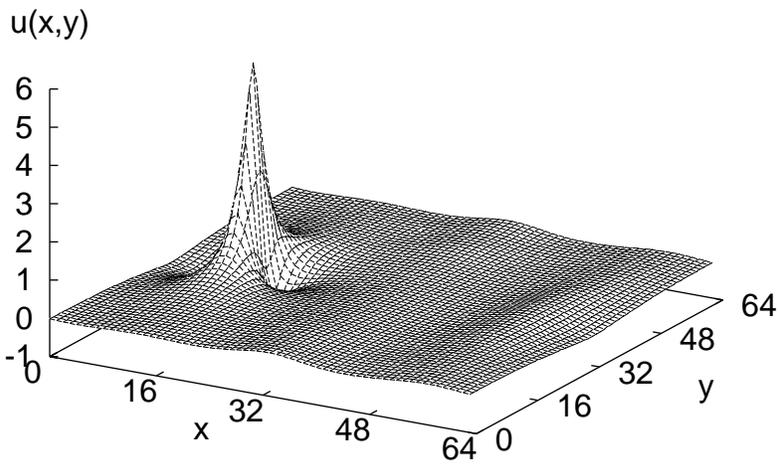
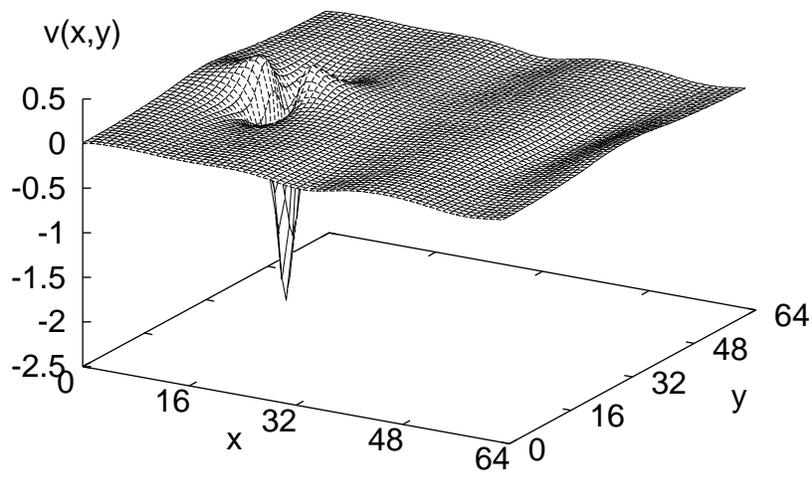
\end{document}